\documentclass[letterpaper, 10 pt, conference]{ieeeconf}  
\IEEEoverridecommandlockouts                              
\usepackage[english]{babel}
\usepackage[T1]{fontenc}

\usepackage[nolist]{acronym}
\usepackage[cmex10]{amsmath}
\usepackage{amssymb}
\usepackage{bm}
\usepackage{booktabs}
\usepackage{color}
\usepackage{float}
\usepackage{graphicx}
\usepackage{hyperref}
\usepackage{import}
\usepackage{mathtools}
\usepackage{multirow}
\usepackage{outline}
\usepackage{paralist}
\usepackage{siunitx}
\usepackage{stmaryrd}
\usepackage{systeme}
\usepackage{units}
\usepackage{url}
\usepackage{tikz}
\usepackage{booktabs}
\usetikzlibrary{tikzmark}

\title{\LARGE \bf Quadrupedal Robotic Guide Dog with Vocal Human-Robot Interaction}
\author{\authorblockN{Kavan Mehrizi}
\authorblockA{Department of Computer Science, Diablo Valley College, Pleasant Hill, CA, 94523, USA}
\authorblockA{kavanmehrizi@berkeley.edu}}

\begin{document}
\maketitle

\begin{abstract}
Guide dogs play a critical role in the lives of many, however training them is a time- and labor-intensive process. We are developing a method to allow an autonomous robot to physically guide humans using direct human-robot communication. The proposed algorithm will be deployed on a Unitree A1 quadrupedal robot and will autonomously navigate the person to their destination while communicating with the person using a speech interface compatible with the robot. This speech interface utilizes cloud based services such as Amazon Polly and Google Cloud to serve as the text-to-speech and speech-to-text engines.
\end{abstract}

\section{Introduction}
The training and maintenance of a traditional guide dog presents challenges to the elderly, frail, and visually-impaired. Each guide dog has to be trained individually in a time and labor intensive process and the skills gained from one dog cannot be implemented into another dog. In addition, guide dogs may get ill or need to retire, which creates a hassle of getting a replacement dog, which may not be a good match for the user \cite{lloyd_investigation_2016}. An autonomous robot that could lead people in need of assistance through a multi-floor building would ease the burdens that come with a traditional guide dog. Most previous robotic guides are bungle-some and are limited to maneuvering in narrow and complex spaces due to their bulky size or rely on physical interaction between the robot and the user, by having them physically hold a leash or rigid arm, without any way for the user to verbally give commands such as to reroute, or stop the robot [2]–[4].\nocite{li_toward_2019} \nocite{wachaja_navigating_2017} \nocite{bruno_development_2019} In addition, none of these guide robots are able to guide and navigate in multi-floor situations. In early 2021, Xiao et al. successfully implemented a robotic quadrupedal robot to guide a subject, however the model relied solely on physical interaction based around a leash and had no way for the person being led to directly communicate to the robot \cite{xiao_robotic_2021}. A small, quadrupedal robot that is both able to directly communicate and listen for commands from the person that is being guided as well as having a leash would solve such issues. We seek to accomplish this by utilizing a Unitree A1 quadrupedal robot \cite{noauthor_unitree_nodate} to autonomously navigate a visually-impaired person in a multi-floor environment by creating algorithms that would allow for a custom wake-up word and communicate with the user via text-to-speech (TTS) and speech-to-text (STT) cloud services.
\label{sec:Introduction}

\section{Methodology}
The robot would be able to vocally communicate with and understand the user using text-to-speech and speech-to-text algorithms. We had to first find basic open source code \cite{noauthor_aws-roboticstts-ros1_2021} that allowed for the integration of Amazon Polly, a cloud service, that allows the robot to speak to the user directly by sending a string of text to Amazon Web Services, which submits that text to Amazon Polly to generate an audio stream. That audio stream is then retrieved from Amazon Polly which is then played through an installed speakerphone on the robot. We then had to make the code compatible with the robot’s infrastructure, which relies on Robot Operating System (ROS). For the robot to understand what the user is saying, we are using Google Cloud and their Speech-to-Text Application Programming Interface (API). Google Speech-to-Text API works by getting audio data from a source, which then runs the audio to convert into a digital line of text. In order to utilize this API, we found open source code from GitHub that is compatible with ROS and configured into the robot’s infrastructure \cite{tan_success_2021}. We gain audio data from the speakerphone on the robot for use with Google Cloud. That string of text is then returned to the STT algorithm, which will look to see if the wake word, which is customizable, has been said. If not, the algorithm ignores whatever was said and will resume to listen. When the wake-up word is said, the string is sent to a word dictionary function that will search for keywords in the resulting text and has preset coordinates based on those keywords. The algorithm then publishes those coordinates to the navigation goal node after understanding where the user wants to go. STT will also publish a string of text to TTS to allow for the robot to respond back to the user. The robot's navigation subscribes to that STT publisher and creates a path to the target point.

\section{Results}
\begin{figure}[h!]
\centering
  \includegraphics[width=0.7\linewidth]{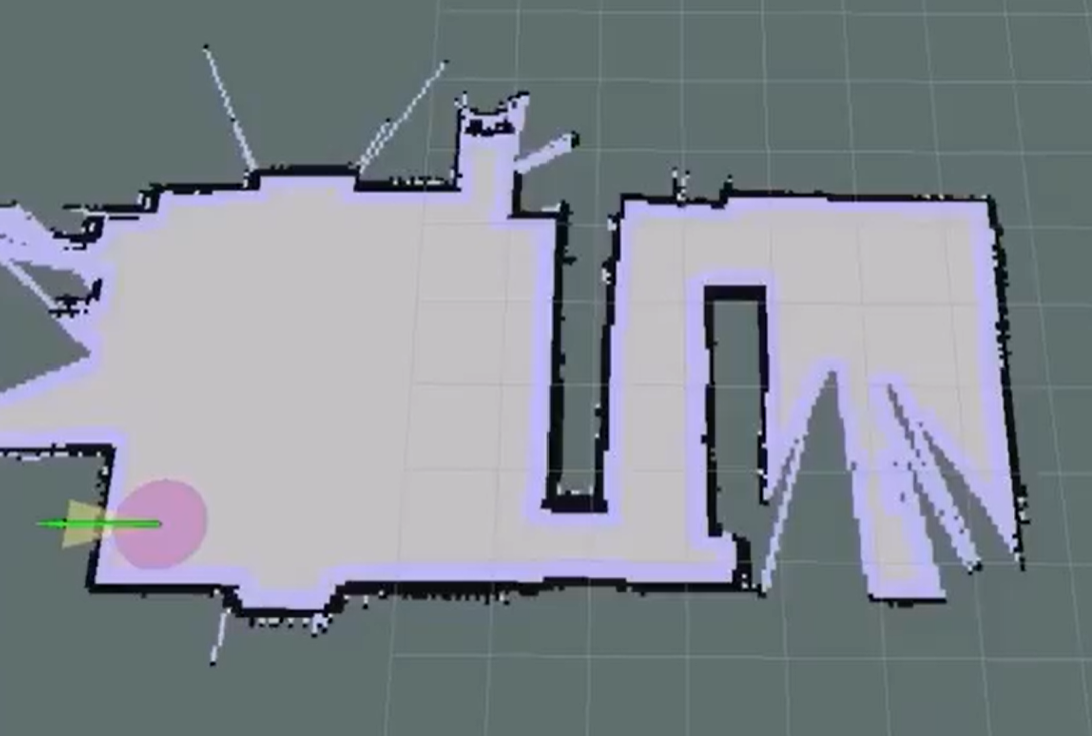}
  \caption{Simulated navigation map showing initial position in the purple circle.}
\label{fig:simulationpic}
  \label{fig:robot1}
\end{figure}

\begin{figure}[h!]
\centering
  \includegraphics[width=0.7\linewidth]{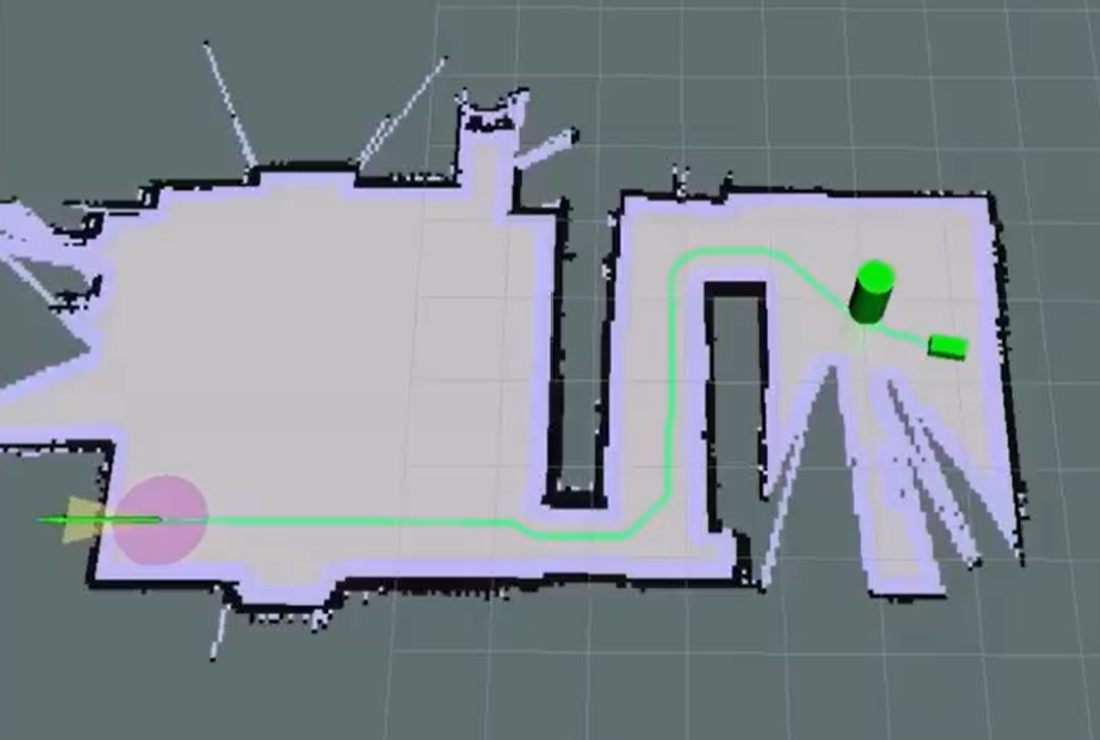}
  \caption{Navigation map after receiving navigation goal coordinates from speech interface. The green figures portray the final position of the user and robot, while the green line is the path created to that final position goal.}
  \label{fig:simulationlab}
\end{figure}

\begin{figure}[h!]
\centering
  \includegraphics[width=0.7\linewidth]{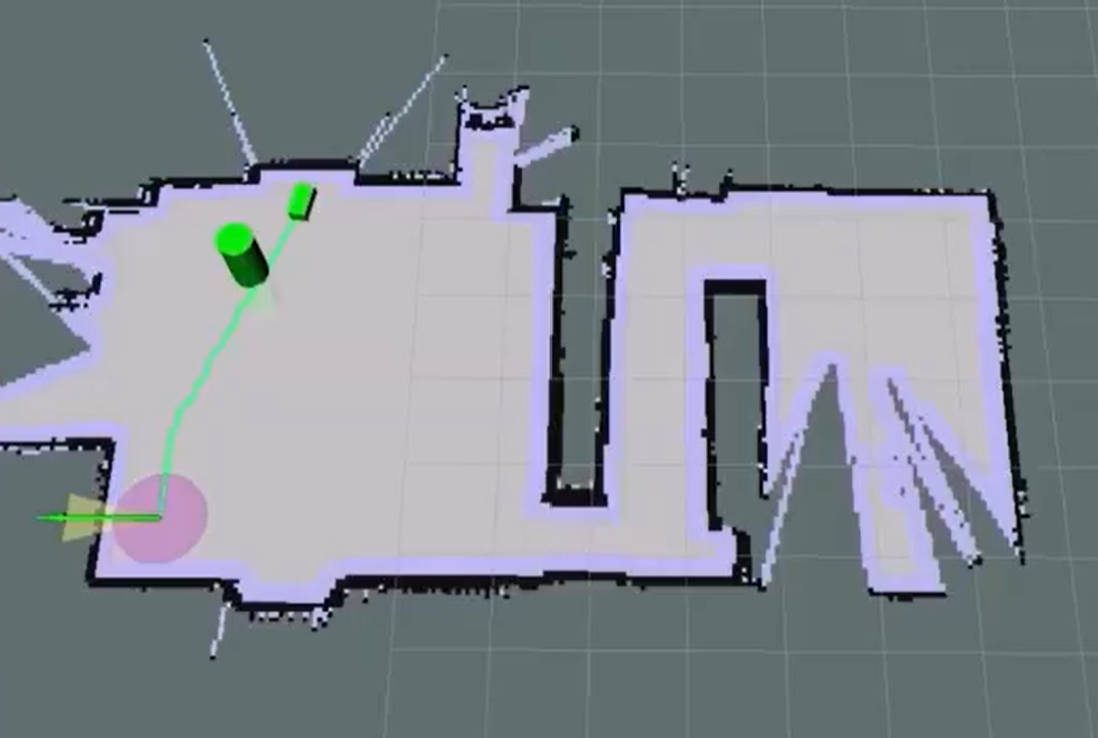}
  \caption{Navigation map after receiving new navigation goal coordinates from speech interface.}
  \label{fig:simulationoffice}
\end{figure}

We tested the speech interface in simulation using a simulated navigation map that the robot would map out using its onboard LiDAR camera shown in Figure \ref{fig:simulationpic}. In this simulation, the user said to the robot, "Hey A1, take me to the lab." The speech interface successfully heard the user’s command and translated the user’s command into a string of text. It then published the pre-set coordinates of the laboratory from the dictionary to the navigation goal node. The robot's navigation was able to subscribe to that node and created a path to that goal location shown in Figure \ref{fig:simulationlab}. Finally the robot responded back to the user saying, "Okay, navigating to the lab." The user then said, "Take me to the office." The speech interface successfully ignored the speech even though it could be a command as the user did not use the wake-up word, which was set to, "Hey A1." The robot's navigation was not affected and no response back was given. It was only when the user said the same sentence but with the wake-up word that the algorithm recognized it as a valid command. This meant that the speech interface sent the coordinates of the office to the robot's navigation pipeline, which resulted in creating a new path shown in Figure \ref{fig:simulationoffice}.

\section{Discussion and Future Work}
These results prove that the TTS and STT engines were able to be integrated with the algorithm created. The algorithm was able to communicate to both the engines and the robot's infrastructure. The robot ignored all irrelevant speech, only sending the string of text to the dictionary function when the wake-up word was said. Unlike previous robots with a speech interface, we are able to have a custom wake-up word and don't rely on an Amazon Echo device \cite{li_toward_2019}. We were able to successfully set a navigation goal solely by verbally communicating a command to the robot. Our previous work, while having a leash, relied on an external computer to input commands, not allowing the user themselves to communicate with the robot \cite{xiao_robotic_2021}. This work improves the user experience by allowing for explicit interaction, not just implicit interaction by the use of a leash.
\par
We currently are further developing the guide dog robot to operate an elevator to allow for multi-floor navigation. In order to facilitate multi-floor navigation, we are currently restructuring the robot’s navigation to take floors into consideration. Having a multi-floor situation means that we need to further develop the speech interface to send coordinates that can relate to what floor level the navigation goal is at. The speech interface will be developed to allow for more commands such as telling the robot to stop at its current position as well as giving the user instructions when needed. We need to further optimize the speech interface such as making it easier to input new commands and new locations into the algorithm.

\section{Conclusion}
Having a speech interface makes it simpler for the user to send commands to the robot. We developed and tested a successful speech interface algorithm that is able to communicate with the TTS and STT engines as well as communicate with the robot's navigation pipeline. The main advantages of this work are that we are able to customize the wake-up word due to having our own proprietary speech interface and are able to create custom commands fairly easily by adding them to the word dictionary. We are able to have a custom wake-up word and are able to integrate this speech interface with a leash while using a maneuverable robot.

\section*{Acknowledgements} 
This work was supported by the Hopper Dean Foundation and National Science Foundation Award \#1757690. The Transfer-to-Excellence Research Experiences for Undergraduates program is sponsored by the National Science Foundation and the Center for Energy Efficient Electronics (NSF \#0939514). I would like to thank my mentor, Zhongyu Li, and my Principal Investigator, Koushil Sreenath, for giving me the opportunity to be apart of their research group. I would also like to thank Nicole McIntyre, Tony Vo Hoang, Sam Mountain, Gary Yang, and the Hybrid Robotics Group for their constant support.


\begin{acronym}
\acro{HP}{high-pass}
\acro{LP}{low-pass}
\end{acronym}

\end{document}